\documentclass[12pt,a4paper]{article}
\usepackage{amsfonts}
\usepackage{amssymb}
\usepackage{amsmath}
\usepackage{latexsym}
\usepackage{graphicx}
\usepackage{hyperref}
\begin{document}

\begin{center}
{\Large \bf On unremovable divergencies in\\ four-dimensional
electrodynamics\\ \vspace{2.3mm}  localized on a domain
wall} \\

\vspace{4mm}

Mikhail N.~Smolyakov\\
\vspace{0.5cm} Skobeltsyn Institute of Nuclear Physics, Moscow
State University,
\\ 119991, Moscow, Russia\\
\end{center}

\begin{abstract}
In this paper a model describing fermions and gauge field
localized on a domain wall is considered. The zero-mode sector of
the model reproduces four-dimensional spinor electrodynamics. It
is shown that in the resulting four-dimensional effective theory
the lowest-order renormalized amplitude of the elastic
$\gamma\gamma$ scattering process is divergent due to the
contributions of higher modes and thus such a scenario can not be
considered as a realistic mechanism for fields localization on a
domain wall.
\end{abstract}

\section{Introduction}
Models with extra dimensions offer many interesting possibilities
for resolving existing problems in theoretical physics. There are
different approaches in such theories, the most popular now being
brane world models. In many brane world models it is assumed that
matter is localized on a brane, embedded into multidimensional
space-time, but very often the particular mechanism of
localization is not considered explicitly. Indeed, the mechanism
of localization of massless fermions is known for a long time, as
it was proposed in \cite{RS}. As for bosons, especially gauge
fields, there are also some ways to localize them. The most known
is the mechanism proposed in \cite{DS} (see also \cite{DR}), where
the ideas of confinement were used to trap the gauge fields.
Another approach is to make the wave functions of massless
four-dimensional gauge bosons constant in extra dimensions (to
ensure the universality of charge, which is important for
constructing realistic models, see \cite{Rubakov}), but
normalizable. The latter can be achieved either by introducing
more than one extra dimension in models with warped geometry
\cite{DRT0}, or by considering an additional overall coupling of
the gauge field to an extra scalar field, for example, dilaton
field \cite{KT} (or just to a weight function depending explicitly
on the extra coordinate as in \cite{ST}), in five-dimensional
models. But it should be necessarily noted that there were earlier
attempts to localize gauge fields on a domain wall, see, for
example, \cite{Barnaveli}.

In this paper we consider a model with both gauge and spinor
fields lowest modes localized on a domain wall. We use the
mechanism of consistent localization of massive fermions on a
domain wall, which is based on the well-known Rubakov-Shaposhnikov
mechanism \cite{RS} and proposed in \cite{DRT} (but we consider it
in the absence of gravity). This mechanism has already been
discussed and utilized in different models
\cite{LT,CG,AAGS,AAGS2}. We present an explicit solution for the
zero fermionic mode, which looks as an ordinary four-dimensional
massive fermion, as well as for the excited modes, which also
represent four-dimensional massive fermions. We also present a
mechanism of localization of gauge vector fields on a domain wall
(again in the absence of gravity), which is motivated by the
results of \cite{KT,ST,Barnaveli} and utilizes the ideas presented
there: an overall coupling of the gauge field to the scalar field
to make the wave function of the gauge field normalizable
\cite{KT} and taking this scalar field to be the one which forms
the domain wall proper \cite{Barnaveli} (it should be noted that
the same idea in connection with localization of gauge fields on
thick branes has been recently utilized in \cite{CHH}). We show
explicitly that both mechanisms lead to quite interesting results:
for bosons, as well as for fermions, there exist nonzero mass gaps
between the lowest (zero) massive or massless localized modes and
the other localized and nonlocalized modes which arise from the
five-dimensional theory. This allows one to consider a combination
of these mechanisms and to construct a simple five-dimensional
model, involving two five-dimensional fermions, interacting with
five-dimensional $U(1)$ gauge field, whose localized sector
represents ordinary four-dimensional spinor electrodynamics. But
it appears that such a theory after a renormalization leads to
infinite amplitudes of some processes involving only the lowest
localized modes in initial and final states (the
$\gamma\gamma\to\gamma\gamma$ scattering is considered, which is
well behaved in ordinary four-dimensional QED). This happens
because of a contribution of the modes from the continuous
spectrum. Thus, such a model can not be considered as realistic
for describing matter fields localization on a brane or a domain
wall and one should look for more consistent scenarios.

\section{Localized massive fermions}
\subsection{Localization mechanism}
The well-known mechanism of localization of fermions is the
Rubakov-Shaposhnikov mechanism \cite{RS}. It describes a
five-dimensional fermion field interacting with a background
scalar field admitting, for example, the kink solution
\begin{equation}\label{kink}
\Phi_{0}=\frac{m}{\sqrt{\lambda}}\tanh\left(\frac{m}{\sqrt{2}}z\right).
\end{equation}
The mechanism allows one to get a localized four-dimensional
chiral mode of the fermion. The problem is that the localized
fermion is massless, whereas it is clear that there should be
localized modes, describing massive four-dimensional fermions, in
order to represent real massive four-dimensional particles.
Indeed, for the kink profile (\ref{kink}) of the scalar field the
left fermion field (left from the four-dimensional point of view)
behaves as
\begin{equation}\label{wfrs}
\Psi_{L}(x,z)\to C_{L}\psi_{L}(x)e^{-\frac{hm}{\sqrt{\lambda}}|z|}
\end{equation}
at large $z$, whereas the right fermion field (right from the
four-dimensional point of view) behaves as
\begin{equation}
\Psi_{R}(x,z)\to C_{R}\psi_{R}(x)e^{\frac{hm}{\sqrt{\lambda}}|z|}
\end{equation}
at large $z$, i.e. its wave function grows at infinity for
$C_{R}\ne 0$ and we should set $C_{R}=0$. Meanwhile, it is
necessary to have both chiralities, preferably with the same wave
functions, in order to describe massive four-dimensional fermions,
so one should introduce two five-dimensional fermion fields. One
fermion field provides the localized left-handed component of the
four-dimensional Dirac spinor, whereas another fermion field
provides the localized right-handed component of the
four-dimensional Dirac spinor. An interaction term of these
fermion fields results in a four-dimensional mass term. To our
knowledge, for the first time this mechanism was utilized in
\cite{DRT} when considering the behavior of matter in brane world
models (i.e. taking into account the gravitational interaction).
Then analogous constructions were used in \cite{LT} for explaining
the origin of the three generations of the Standard Model fermions
and in \cite{AAGS,AAGS2} for constructing domain walls generated
by a fermion self-interaction. Below we will recall this mechanism
and examine some of its properties. We will show how it provides
the existence of four-dimensional massive fermions localized on a
domain wall.

Let us take the five-dimensional action corresponding to the model
of \cite{DRT}, but in the absence of gravity (see also the simple
model discussed in the beginning of \cite{AAGS}). In our notations
this action has the form:
\begin{eqnarray}\label{fermact}
S=\int
\left[i\bar\Psi_{1}\Gamma^{M}\partial_{M}\Psi_{1}-h\Phi\bar\Psi_{1}\Psi_{1}
+i\bar\Psi_{2}\Gamma^{M}\partial_{M}\Psi_{2}+h\Phi\bar\Psi_{2}\Psi_{2}-\right.\\
\nonumber\left.-M\left(\bar\Psi_{1}\Psi_{2}+\bar\Psi_{2}\Psi_{1}\right)\right]d^{4}xdz,
\end{eqnarray}
where $\Psi_{1}$ and $\Psi_{2}$ are different five-dimensional
fermion fields, $h>0$, $M=0,..,4$, $\Gamma^{\mu}=\gamma^{\mu}$,
$\Gamma^{4}=i\gamma^{5}$. The equations of motion coming from this
action have the form
\begin{eqnarray}\label{eq1}
i\gamma^{\mu}\partial_{\mu}\Psi_{1}-\gamma^{5}\Psi'_{1}-h\Phi_{0}\Psi_{1}-M\Psi_{2}=0,
\\\label{eq2}
i\gamma^{\mu}\partial_{\mu}\Psi_{2}-\gamma^{5}\Psi'_{2}+h\Phi_{0}\Psi_{2}-M\Psi_{1}=0,
\end{eqnarray}
where $'$ denotes $\partial_{4}$ and $\Phi_{0}(z)$ is a domain
wall profile (see equations (1.9) and the subsequent discussion in
\cite{AAGS}). Let us take the following form of the solution to
these equations:
\begin{eqnarray}\label{anz1}
&&\Psi_{1}(x,z)=f(z)\psi_{L}(x),\qquad
\gamma^{5}\psi_{L}=\psi_{L},\\\label{anz2}
&&\Psi_{2}(x,z)=f(z)\psi_{R}(x),\qquad
\gamma^{5}\psi_{R}=-\psi_{R}.
\end{eqnarray}
Substituting (\ref{anz1}), (\ref{anz2}) into (\ref{eq1}),
(\ref{eq2}) we easily get:
\begin{eqnarray}\label{sol1}
&&\Psi_{1}(x,z)=Ce^{-h\int_{0}^{z}\Phi_{0}(z')dz'}\psi_{L}(x),\qquad
i\gamma^{\mu}\partial_{\mu}\psi_{L}-M\psi_{R}=0,\\\label{sol2}
&&\Psi_{2}(x,z)=Ce^{-h\int_{0}^{z}\Phi_{0}(z')dz'}\psi_{R}(x),\qquad
i\gamma^{\mu}\partial_{\mu}\psi_{R}-M\psi_{L}=0,
\end{eqnarray}
where $C$ is a normalization constant. The four-dimensional
equations in (\ref{sol1}) and (\ref{sol2}) can be combined into
one equation by taking
$$\psi_{L}(x)+\psi_{R}(x)=\psi(x)$$ and we get
\begin{eqnarray}\label{sol3}
i\gamma^{\mu}\partial_{\mu}\psi-M\psi=0.
\end{eqnarray}
The latter equation is nothing but the Dirac equation for a
massive fermion. Thus, we get the localized mode, which describes
a massive four-dimensional fermion. This massive four-dimensional
fermion appears to be the superposition of the zero modes of the
two initial five-dimensional fermions.

The corresponding four-dimensional effective action for this mode
can be easily obtained and has the form
\begin{eqnarray}
S=C^{2}\int e^{-2h\int_{0}^{z}\Phi_{0}(z')dz'} dz\int
\left(i\bar\psi\gamma^{\mu}\partial_{\mu}\psi-M\bar\psi\psi\right)d^{4}x.
\end{eqnarray}
If $C$ is chosen in a proper way, we get a canonically-normalized
four-dimensional effective action having the standard form
\begin{eqnarray}
S_{eff}=\int
\left(i\bar\psi\gamma^{\mu}\partial_{\mu}\psi-M\bar\psi\psi\right)d^{4}x.
\end{eqnarray}
Of course, this mechanism, as well as the original mechanism
\cite{RS}, can be utilized not only with the kink profile of the
background scalar field, but with any background scalar field
profile suitable for obtaining normalizable wave functions of
fermions.

\subsection{Mass gap and excited fermion states}
Now let us turn to examining the properties of this mechanism.
Indeed, it is not evident that the localized mode presented in the
previous section is the lowest one and that there is a nonzero
mass gap between the lowest mode and the next possible localized
mode or the continuous spectrum. To this end let us obtain the
second order differential equations for the spinors $\Psi_{1}$ and
$\Psi_{2}$. This can be done by the standard technique and we
obtain
\begin{eqnarray}\label{ms1}
-\partial^{\mu}\partial_{\mu}\Psi_{1}-M^{2}\Psi_{1}&=&-\Psi''_{1}-
h\Phi_{0}'\gamma^{5}\Psi_{1}+h^{2}\Phi_{0}^{2}\Psi_{1},\\
\label{ms2}
-\partial^{\mu}\partial_{\mu}\Psi_{2}-M^{2}\Psi_{2}&=&-\Psi''_{2}+
h\Phi_{0}'\gamma^{5}\Psi_{2}+h^{2}\Phi_{0}^{2}\Psi_{2}.
\end{eqnarray}
For $M=0$ the first of these equations has exactly the same form
as the one which can be obtained in the model \cite{RS} with just
one five-dimensional spinor. This is enough to make some
statements about other possible localized modes -- we know that in
the model \cite{RS} there may be (or may not be) other localized
states with larger masses and that there is a nonzero mass gap
between the lowest mode and the continuous spectrum. In the model
discussed here we can also easily obtain analogous estimates. To
this end one should solve the equation for the wave function,
coming from (\ref{ms1}) (equation (\ref{ms2}) also gives the same
equation):
\begin{equation}\label{eqgenferm}
\left(\mu^2-M^{2}\right)f(z)=-f''(z)\mp
h\Phi_{0}'f(z)+h^{2}\Phi_{0}^{2}f(z),
\end{equation}
where $\mp$ comes from $\gamma^{5}$ acting on the left-handed and
right-handed fermions respectively. For the kink profile
(\ref{kink}) of the background field we get
\begin{equation}\label{eqaux}
\left(\mu^2-M^{2}-h^{2}\frac{m^{2}}{\lambda}\right)f(z)=-f''(z)-
\frac{h^{2}\frac{m^{2}}{\lambda}\pm
h\frac{m^{2}}{\sqrt{2\lambda}}}{\cosh^{2}\left(\frac{m}{\sqrt{2}}z\right)}f(z).
\end{equation}
It is the standard Schr\"odinger-type equation with the known
solution -- the eigenvalues, corresponding to the localized modes,
are (see, for example, \cite{Flugge})
\begin{equation}\label{nummodes}
\mu^2=M^{2}+\frac{\sqrt{2}hm^{2}}{\sqrt{\lambda}}\left(n+\frac{1}{2}\mp\frac{1}{2}\right)-
\frac{m^{2}}{2}\left(n+\frac{1}{2}\mp\frac{1}{2}\right)^{2},\qquad
n\le\frac{\sqrt{2}h}{\sqrt{\lambda}}-\frac{1}{2}\pm\frac{1}{2}.
\end{equation}
Here $n=0,1,2,...$ and we suppose that
$\frac{\sqrt{2}h}{\sqrt{\lambda}}\ge 1$. The lowest mode, as
expected, has the mass $\mu_{0}=M$. The next eigenvalue is
$$\mu_{1}=\sqrt{M^{2}+\frac{m^{2}}{\sqrt{2}}\left(\frac{2h}{\sqrt{\lambda}}-
\frac{1}{\sqrt{2}}\right)}.$$ Thus, the mass gap is at least
\begin{equation}\label{massgap}
\Delta\mu=\sqrt{M^{2}+\frac{m^{2}}{\sqrt{2}}\left(\frac{2h}{\sqrt{\lambda}}-
\frac{1}{\sqrt{2}}\right)}-M.
\end{equation}

Now let us show how the excited modes look like in the scenario
under consideration. Let us consider the $n$-th Kaluza-Klein
level, $n\ge 1$, and suppose that it contains particles with the
same four-dimensional mass $\mu$. It follows from
(\ref{eqgenferm}) that the wave function $f(z)$ of the fermions,
for which
$\gamma^{5}\Psi_{1}=\Psi_{1},\,\gamma^{5}\Psi_{2}=-\Psi_{2}$ hold,
fulfills the equation
\begin{equation}\label{e2}
\left(\mu^2-M^{2}\right)f(z)=-f''(z)-
h\Phi_{0}'f(z)+h^{2}\Phi_{0}^{2}f(z),
\end{equation}
whereas the wave function $\tilde f(z)$ of the fermions, for which
$\gamma^{5}\Psi_{1}=-\Psi_{1},\,\gamma^{5}\Psi_{2}=\Psi_{2}$ hold,
fulfills the equation
\begin{equation}\label{e3}
\left(\mu^2-M^{2}\right)\tilde f(z)=-{\tilde f}''(z)+
h\Phi_{0}'\tilde f(z)+h^{2}\Phi_{0}^{2}\tilde f(z).
\end{equation}
It is not difficult to show that equations (\ref{e2}) and
(\ref{e3}) can be obtained from the following system of equations
\begin{eqnarray}\label{e4}
&f'+h\Phi_{0} f=(\mu+M){\tilde f},\\
\label{e5} &{\tilde f}'-h\Phi_{0}{\tilde f}=-(\mu-M)f,
\end{eqnarray}
which will be used below. Equations (\ref{e2}) and (\ref{e3})
suggest the decomposition for the five-dimensional spinors
$\Psi_{1}$ and $\Psi_{2}$:
\begin{eqnarray}\label{e6}
\Psi_{1}(x,z)=f(z)\psi_{L}(x)+\frac{\tilde f(z)}{D}\hat\psi_{R}(x),\\
\label{e7} \Psi_{2}(x,z)=f(z)\psi_{R}(x)-\frac{\tilde
f(z)}{D}\hat\psi_{L}(x),
\end{eqnarray}
where $D^{2}=\int {\tilde f}^{2}dz$,
$\psi_{L}=\gamma^{5}\psi_{L}$, $\psi_{R}=-\gamma^{5}\psi_{R}$,
$\hat\psi_{L}=\gamma^{5}\hat\psi_{L}$,
$\hat\psi_{R}=-\gamma^{5}\hat\psi_{R}$. We also impose the
normalization condition $\int f^2 dz=1$, which results in
$D^{2}=\frac{\mu-M}{\mu+M}$ (the value of $D$ can be obtained from
equations (\ref{e4}), (\ref{e5}).

Substituting decomposition (\ref{e6}), (\ref{e7}) into action
(\ref{fermact}), taking into account (\ref{e4}), (\ref{e5}) and
then integrating over the coordinate of the extra dimension, we
get the effective four-dimensional action
\begin{eqnarray}
S_{eff}=\int d^{4}x \left(
i\bar\psi\gamma^{\mu}\partial_{\mu}\psi+i\bar{\hat\psi}\gamma^{\mu}\partial_{\mu}\hat\psi-
M(\bar\psi\psi-\bar{\hat\psi}\hat\psi)-\sqrt{\mu^{2}-M^2}\left(\bar\psi\hat\psi+
\bar{\hat\psi}\psi\right)\right),
\end{eqnarray}
where $\psi=\psi_{L}+\psi_{R}$,
$\hat\psi=\hat\psi_{L}+\hat\psi_{R}$. The mass matrix can be
diagonalized with the help of the rotation
\begin{eqnarray}\label{rotation1}
\psi(x)&=&\psi_{1}(x)\cos(\theta)+\psi_{2}(x)\sin(\theta),\\
\label{rotation2}
\hat\psi(x)&=&\psi_{1}(x)\sin(\theta)-\psi_{2}(x)\cos(\theta),
\end{eqnarray}
where $\tan(2\theta)=\frac{\sqrt{\mu^{2}-M^2}}{M}$. The last step
is to make the redefinition $\psi_{2}\to \gamma^{5}\psi_{2}$ in
order to get the conventional sign of the mass term of the
four-dimensional fermion $\psi_{2}$. Finally, we get
\begin{eqnarray}
S_{eff}=\int d^{4}x
\sum\limits_{i=1}^{2}\left[i\bar\psi_{i}\gamma^{\mu}\partial_{\mu}\psi_{i}-
\mu\bar{\psi}_{i}\psi_{i}\right].
\end{eqnarray}

The equation for the function $f$ for the kink profile of the
background scalar field looks like
\begin{equation}\label{eqauxexc}
\left(\mu^2-M^{2}-h^{2}\frac{m^{2}}{\lambda}\right)f(z)=-f''(z)-
\frac{h^{2}\frac{m^{2}}{\lambda}+
h\frac{m^{2}}{\sqrt{2\lambda}}}{\cosh^{2}\left(\frac{m}{\sqrt{2}}z\right)}f(z).
\end{equation}
We know one solution to this equation -- it is simply given by
(\ref{sol1}), (\ref{sol2}), $\mu=M$ in this case and from
(\ref{e4}) we get ${\tilde f}\equiv 0$. Indeed, this is the lowest
localized mode.

The next localized solution to equation (\ref{eqauxexc}) has the
form
\begin{equation}\label{firstloc}
f_{1}=C\frac{\sinh\left(\frac{m}{\sqrt{2}}z\right)}{\cosh^{\alpha}\left(\frac{m}{\sqrt{2}}z\right)},
\end{equation}
where $\alpha=\frac{h\sqrt{2}}{\sqrt{\lambda}}$ and $C$ is the
normalization constant. The four-dimensional mass of this mode is
$\mu_{1}^{2}=M^{2}+m^{2}\left(\frac{\sqrt{2}h}{\sqrt{\lambda}}-\frac{1}{2}\right)$.
From (\ref{e4}) we get
\begin{equation}\label{firstloctilde}
\tilde
f_{1}=C\frac{m}{\sqrt{2}(\mu_{1}+M)\cosh^{\alpha-1}\left(\frac{m}{\sqrt{2}}z\right)}.
\end{equation}
Of course, there can be other localized modes, their number $N$ is
defined by (see (\ref{nummodes}))
\begin{equation}\label{locNumber}
N=\textrm{max}\left\{n\in\mathbb{Z}\left|\right.0<n<1+\frac{\sqrt{2}h}{\sqrt{\lambda}}\right\}.
\end{equation}
They are solutions to equations (\ref{eqauxexc}) and (\ref{e4}).

With the help of (\ref{e4}) and (\ref{e5}) it is not difficult to
show that the following orthogonality conditions are fulfilled in
the general case
\begin{equation}
\int f_{n}f_{k} dz=0,\qquad \int{\tilde f_{n}}{\tilde f_{k}}dz=0,
\qquad n\ne k.
\end{equation}
This is also valid for the modes from the continuous spectrum.

It should be noted that in general there exist two linearly
independent solutions to equation (\ref{e2}) corresponding to each
eigenvalue $\mu$ (the same is valid for equation (\ref{e3}) also).
For the modes from the discrete spectrum one solution is
normalizable, whereas another one grows at infinity and should be
set to zero. Thus, in this case only one solution to each equation
survives. But for the modes from the continuous spectrum both
solutions to equation (\ref{e2}) (as well as to equation
(\ref{e3})) have a good behavior at infinity in the general case
and thus both solutions should be taken into account (one can
choose, say, odd and even solutions if $\Phi_{0}(z)$ is odd). We
will use the notations $f_{(\mu,j)}$ and $\tilde f_{(\mu,j)}$,
$j=1,2$, for these solutions to equations (\ref{e2}) and
(\ref{e3}).

Thus, for the localized modes we can set
\begin{equation}\label{normlocm}
\int f_{n}f_{k} dz=\delta_{nk},
\end{equation}
whereas for the modes from the continuous spectrum we set
\begin{equation}
\int f_{(\mu,j)}f_{(\mu',j')}dz=\delta(\mu-\mu')\delta_{jj'}.
\end{equation}
Now let us obtain the full four-dimensional effective action for
the fermions. To this end we substitute
\begin{eqnarray}
\Psi_{1}&=&\sum\limits_{n=0}^{N}\left(f_n(z)\psi_{L}^n(x)+\frac{{\tilde
f_n(z)}}{D_{n}}\hat\psi_{R}^n(x)\right)+\\
\nonumber &+&\sum\limits_{j=1}^{2}\int
d\mu\left(f_{(\mu,j)}(z)\psi_{L}^{(\mu,j)}(x)+\frac{{\tilde
f_{(\mu,j)}(z)}}{D_{(\mu)}}\hat\psi_{R}^{(\mu,j)}(x)\right),\\
\Psi_{2}&=&\sum\limits_{n=0}^{N}\left(f_n(z)\psi_{R}^n(x)-\frac{{\tilde
f_n(z)}}{D_{n}}\hat\psi_{L}^n(x)\right)+\\
\nonumber &+&\sum\limits_{j=1}^{2}\int
d\mu\left(f_{(\mu,j)}(z)\psi_{R}^{(\mu,j)}(x)-\frac{{\tilde
f_{(\mu,j)}(z)}}{D_{(\mu)}}\hat\psi_{L}^{(\mu,j)}(x)\right),
\end{eqnarray}
where $\tilde f_{0}\equiv 0$, into (\ref{fermact}) and after some
algebra (including the rotation (\ref{rotation1}),
(\ref{rotation2}) with
$\tan(2\theta_{n})=\frac{\sqrt{\mu_{n}^{2}-M^2}}{M}$, $n\ge 1$,
$\tan(2\theta_{(\mu,j)})=\frac{\sqrt{\mu^{2}-M^2}}{M}$ at each
Kaluza-Klein level) we get the four-dimensional action
\begin{eqnarray}
S_{eff}=\int \left(i\bar\psi\gamma^{\nu}\partial_{\nu}\psi
-M\bar\psi\psi\right)d^{4}x+\\
\nonumber +\sum\limits_{i=1}^{2}\left(\sum\limits_{n=1}^{N}\int
\left(i\bar\psi^{n}_{i}\gamma^{\nu}\partial_{\nu}\psi^{n}_{i}
-\mu_{n}\bar\psi^{n}_{i}\psi^{n}_{i}\right)d^{4}x+\right.\\
\nonumber \left.+\sum\limits_{j=1}^{2}\int d\mu\int
\left(i\bar\psi^{(\mu,j)}_{i}\gamma^{\nu}\partial_{\nu}\psi^{(\mu,j)}_{i}
-\mu\bar\psi^{(\mu,j)}_{i}\psi^{(\mu,j)}_{i}\right)d^{4}x\right),
\end{eqnarray}
where $\psi$, $\psi^{n}_{i}$ and $\psi^{(\mu,j)}_{i}$ are the
four-component spinors. We see that each Kaluza-Klein level,
except the lowest one with $\mu_{0}=M$, contains two (discrete
spectrum) or four (continuous spectrum) four-dimensional fermions
with the equal masses $\mu$. The doubling parameterized by $i$ is
of the same nature as that in the UED models, see, for example,
\cite{UED}.

We have shown that for the kink profile of the background scalar
field the excited localized modes really exist and correspond to
four-dimensional massive fermions. The continuous spectrum, as it
can be seen from (\ref{eqaux}), starts at
$$
\mu=\sqrt{M^{2}+\frac{h^{2}}{\lambda}m^{2}}.
$$

Quite an interesting case is that with
$\frac{h}{\sqrt{\lambda}}=\frac{1}{\sqrt{2}}$. In this case
$\alpha=1$ in eqs. (\ref{firstloc}), (\ref{firstloctilde}) and,
according to (\ref{locNumber}), there is only one localized mode
with the mass $\mu_{0}=M$, the continuous spectrum starts at
$\mu=\sqrt{M^{2}+\frac{m^{2}}{2}}$. Thus, for the mass gap we have
\begin{equation}
\Delta\mu=\sqrt{M^{2}+\frac{m^{2}}{2}}-M
\end{equation}
and for $m\gg M$
\begin{equation}
\Delta\mu\approx\frac{m}{\sqrt{2}}.
\end{equation}
Below for simplicity we will study this case
$\frac{h}{\sqrt{\lambda}}=\frac{1}{\sqrt{2}}$.

Thus, we see that the model proposed in \cite{DRT} provides us
massive fermions, localized on a domain wall, and a nonzero mass
gap between these modes and the subsequent continuous spectrum.

\section{Localized U(1) gauge field}
Now we turn to the gauge fields. We consider the simplest
five-dimensional $U(1)$ gauge field. The corresponding
five-dimensional action is chosen to be
\begin{equation}\label{gauge1}
S=-\int\xi^{2}\left(\Phi_{v}^2-\Phi^2\right)\frac{1}{4}F^{MN}F_{MN}d^{4}xdz,
\end{equation}
where $F_{MN}=\partial_{M}A_{N}-\partial_{N}A_{M}$,
$M,N=0,1,..,4$, here and below the parameter $\xi$ is chosen so
that the dimension of $A_{M}$ is mass (like in a four-dimensional
theory). We also suppose that the background solution is such that
$A_{M}\equiv 0$, $\Phi(x,z)=\Phi_{0}(z)$ represents the domain
wall, $\Phi\to\pm\Phi_{v}$ at $z\to\pm\infty$ and
$\Phi_{v}^2-\Phi^2>0$.

Action (\ref{gauge1}) looks rather nonstandard because of the term
$\Phi_{v}^2-\Phi^2$. Nevertheless, an analogous form of
interaction (namely, $(\Phi_{v}^2-\Phi^2)^2$) was discussed in
\cite{Barnaveli} while constructing a mechanism of gauge field
trapping on a domain wall and in \cite{Barnaveli0} (see also
\cite{{CHH}}).

From the very beginning it is convenient to work in the gauge
$A_{4}\equiv 0$. After imposing this gauge we are left with
residual gauge transformations $A_{\mu}\to
A_{\mu}-\partial_{\mu}\alpha$ with $\alpha(x,z)=\alpha(x)$.

Let us represent the field $A_{\mu}$ as
\begin{equation}\label{gauge2}
A_{\mu}(x,z)=f(z)a_{\mu}(x),
\end{equation}
where
\begin{equation}\label{gauge3}
\partial^{\mu}f_{\mu\nu}+\mu^2 a_{\nu}=0,\qquad
f_{\mu\nu}=\partial_{\mu}a_{\nu}-\partial_{\nu}a_{\mu}.
\end{equation}
In this case equations of motion coming from (\ref{gauge1}) take
the form
\begin{eqnarray}\label{gauge4}
\mu^2f+f''-\frac{2\Phi_{0}\Phi_{0}'}{\Phi_{v}^2-\Phi_{0}^2}f'=0,\\
\label{gauge5} f'\partial^{\mu}a_{\mu}=0.
\end{eqnarray}
We see that there exists the lowest massless ($\mu=0$) mode
\begin{eqnarray}\label{zeroa}
A_{\mu}(x,z)=Ca_{\mu}(x),
\end{eqnarray}
where $C$ is a constant, satisfying Maxwell's equations
\begin{eqnarray}
\partial^{\mu}f_{\mu\nu}=0.
\end{eqnarray}
There can also exist massive modes with $f(z)\ne\textrm{const}$,
$\mu\ne 0$, in this case equation (\ref{gauge5}) results in an
additional constraint
\begin{equation}
\partial^{\mu}a_{\mu}=0,
\end{equation}
but this constraint is automatically satisfied in the case of
four-dimensional massive vector field.

Now let us examine other possible solutions to equation
(\ref{gauge4}). Let us make a substitution
\begin{equation}\label{subst1}
f=\frac{1}{\left(\Phi_{v}^2-\Phi_{0}^2\right)^{\frac{1}{2}}}y(z).
\end{equation}
Then we can get an equation
\begin{equation}\label{eqscal3}
\mu^{2}y=-y''-\frac{(\Phi_{0}''\Phi_{0}+{\Phi_{0}'}^2)(\Phi_{v}^2-\Phi_{0}^2)+{\Phi_{0}'}^2\Phi_{0}^2}{(\Phi_{v}^2-\Phi_{0}^2)^2}y,
\end{equation}
which again has the standard form of the quantum mechanical
Schr\"odinger equation. Let there exist solutions $y_{n}(z)$ to
this equation, corresponding to eigenvalues $\mu_{n}^{2}$.
Substituting (\ref{subst1}) into (\ref{gauge1}) and using
(\ref{eqscal3}) we get a four-dimensional effective action
\begin{equation}\label{act2}
S_{eff}=\sum_{n}\int\xi^{2}
y_{n}^{2}(z)dz\int\left(-\frac{1}{4}f_{\mu\nu}^n
f^{\mu\nu}_n+\frac{\mu_{n}^{2}}{2}a_{\mu}^n
a^{\mu}_n\right)d^{4}x.
\end{equation}
It is worth mentioning that there can also exist a continuous
spectrum, which should be taken into account in (\ref{act2}).
There also always exists the mode
$y_{0}=C(\Phi_{v}^2-\Phi_{0}^2)^{\frac{1}{2}}$, $\mu_{0}=0$ (see
(\ref{zeroa})). The eigenfunctions $y_{n}(z)$ are orthogonal with
respect to
\begin{equation}
\int y_{n}y_{k} dz=0,\qquad n\ne k.
\end{equation}
Thus, for localized modes we set $\int \xi^{2}y_{n}^{2}(z)dz=1$,
for modes from the continuous spectrum we set $\int
\xi^{2}y_{(\mu,j)}(z)y_{(\mu',j')}(z)dz=\delta(\mu-\mu')\delta_{jj'}$
and get the four-dimensional effective action
\begin{equation}\label{act3}
S_{eff}=\sum_{n}\int\left(-\frac{1}{4}f_{\rho\nu}^n
f^{\rho\nu}_n+\frac{\mu_{n}^{2}}{2}a_{\rho}^n
a^{\rho}_n\right)d^{4}x+\sum\limits_{j=1}^{2}\int
d\mu\int\left(-\frac{1}{4}f_{\rho\nu}^{(\mu,j)}
f^{\rho\nu}_{(\mu,j)}+\frac{\mu^{2}}{2}a_{\rho}^{(\mu,j)}
a^{\rho}_{(\mu,j)}\right)d^{4}x.
\end{equation}
The key feature of the mechanism presented above is that although
there exists the mode whose wave function is constant in extra
dimension, we have an effective four-dimensional action for this
mode. This property of the mechanism appears to be very useful
when considering non-abelian gauge fields, which, of course, can
be localized in the same way.

For the kink background (\ref{kink}) equation (\ref{eqscal3})
takes the form (in this case $\Phi_{v}=\frac{m}{\sqrt{\lambda}}$)
\begin{equation}\label{eqscal4}
\left(\mu^{2}-\frac{m^2}{2}\right)y=-y''-\frac{m^{2}}{\cosh^2\left(\frac{m}{\sqrt{2}}z\right)}y
\end{equation}
(compare with eq. (\ref{eqauxexc})). For the given set of the
parameters (it corresponds to
$\frac{h}{\sqrt{\lambda}}=\frac{1}{\sqrt{2}}$ in (\ref{eqauxexc}))
there exists one normalizable mode with
\begin{equation}
\mu_{0}=0.
\end{equation}
The continuous spectrum starts at $\mu=\frac{m}{\sqrt{2}}$. Thus,
we get the localized mode with the zero four-dimensional mass and
the nonzero mass gap, defined by the domain wall thickness $1/m$,
between this mode and the continuous spectrum. The effective
action of the zero mode has the form
\begin{equation}\label{sccoup0}
S_{eff}^0=-\int\xi^{2}\left(\Phi_{v}^2-\Phi_{0}^2\right)f_{0}^{2}(z)dz\int\frac{1}{4}f^{\mu\nu}f_{\mu\nu}d^{4}x.
\end{equation}
For the properly normalized $f_{0}(x)$ we get
\begin{equation}\label{effactgauge1}
S_{eff}^0=-\int\frac{1}{4}f^{\mu\nu}f_{\mu\nu}d^{4}x
\end{equation}
and
\begin{equation}\label{zeromode}
A^{0}_{\mu}(x,z)=\frac{\sqrt{\lambda}}{2^{\frac{3}{4}}\xi\sqrt{m}}a_{\mu}(x).
\end{equation}
It is necessary to note that there exist residual gauge
transformations $a_{\mu}(x)\to
a_{\mu}(x)-\partial_{\mu}\tilde\alpha(x)$. Thus, as expected, the
localized field $a_{\mu}$ represents four-dimensional $U(1)$ gauge
field. The constant wave function of the zero mode in
(\ref{zeromode}) clearly ensures the universality of charge and
allows one to consider non-abelian gauge fields.

As it was noted above, in some sense this mechanism of gauge field
localization is similar to the "gravitational localization"
mechanism in models with more than one extra dimension
\cite{DRT0}, where the constant wave function of the vector zero
mode also exists and this mode is normalizable, and to the
mechanism of \cite{KT} with analogous consequences. But, contrary
to the case of \cite{DRT0}, in our case there exists a nonzero
mass gap between the localized zero mode and the continuous
spectrum.

\section{Effective four-dimensional electrodynamics}
As an example, it is interesting to consider the simplest theory,
involving gauge bosons and fermions, which can be easily localized
on the brane -- electrodynamics. We take the following
five-dimensional action:
\begin{eqnarray}\label{sact}
S=\int
\left[i\bar\Psi_{1}\Gamma^{M}\left(\partial_{M}-ieA_{M}\right)\Psi_{1}-h\Phi\bar\Psi_{1}\Psi_{1}
+i\bar\Psi_{2}\Gamma^{M}\left(\partial_{M}-ieA_{M}\right)\Psi_{2}+h\Phi\bar\Psi_{2}\Psi_{2}-\right.\\
\nonumber\left.-M\left(\bar\Psi_{1}\Psi_{2}+\bar\Psi_{2}\Psi_{1}\right)-\frac{1}{4}\xi^{2}\left(\Phi_{v}^2-
\Phi^2\right)F^{MN}F_{MN}\right]d^{4}xdz.
\end{eqnarray}
We also take the kink profile (\ref{kink}) for the background
scalar field, $\frac{h}{\sqrt{\lambda}}=\frac{1}{\sqrt{2}}$ and
impose the gauge $A_{4}\equiv 0$. The spinor and vector fields can
be represented as
\begin{eqnarray}\label{sanz1}
A_{\nu}(x,z)=\frac{\sqrt{\lambda}}{2^{\frac{3}{4}}\xi\sqrt{m}}a_{\nu}(x)+\frac{\sqrt{\lambda}}{m}\cosh\left(\frac{m}{\sqrt{2}}z\right)
\sum\limits_{j=1}^{2}\int\limits_{\frac{m}{\sqrt{2}}}^{\infty}d\mu\, y_{(\mu,j)}(z)a_{\nu}^{(\mu,j)}(x) ,\\
\label{sanz2}\Psi_{1}(x,z)=\frac{\sqrt{m}}{2^{\frac{3}{4}}\cosh\left(\frac{m}
{\sqrt{2}}z\right)}\psi_{L}(x)+\sum\limits_{j=1}^{2}\int\limits_{\sqrt{\frac{m^2}{2}+M^{2}}}^{\infty}
d\mu\left(f_{(\mu,j)}(z)\psi_{L}^{(\mu,j)}(x)+\frac{{\tilde
f_{(\mu,j)}(z)}}{D_{(\mu)}}\hat\psi_{R}^{(\mu,j)}(x)\right),\\
\label{sanz3}
\Psi_{2}(x,z)=\frac{\sqrt{m}}{2^{\frac{3}{4}}\cosh\left(\frac{m}
{\sqrt{2}}z\right)}\psi_{R}(x)+\sum\limits_{j=1}^{2}\int\limits_{\sqrt{\frac{m^2}{2}+M^{2}}}^{\infty}d\mu\left(f_{(\mu,j)}(z)
\psi_{R}^{(\mu,j)}(x)-\frac{{\tilde
f_{(\mu,j)}(z)}}{D_{(\mu)}}\hat\psi_{L}^{(\mu,j)}(x)\right).
\end{eqnarray}
Substituting (\ref{sanz1})--(\ref{sanz3}) into (\ref{sact}) and
performing calculations analogous to those presented in
Section~2.2, we get a four-dimensional effective action
\begin{eqnarray}\label{eqspact}
S_{eff}=\int d^{4}x
\biggl[i\bar\psi\gamma^{\nu}\left(\partial_{\nu}-ie_{4}a_{\nu}\right)\psi-M\bar\psi\psi-\frac{1}{4}
f^{\rho\nu}f_{\rho\nu}+\\ \nonumber
+\sum\limits_{j=1}^{2}\int\limits_{\sqrt{\frac{m^2}{2}+M^{2}}}^{\infty}d\mu\Biggl(\sum\limits_{i=1}^{2}\left(i\bar\psi^{(\mu,j)}_{i}\gamma^{\nu}\left(\partial_{\nu}
-ie_{4}a_{\nu}\right)\psi^{(\mu,j)}_{i}
-\mu\bar\psi^{(\mu,j)}_{i}\psi^{(\mu,j)}_{i}\right)+\\ \nonumber +
C_{\psi,a}^{(\mu)}a_{\nu}^{(\tilde\mu,j)}\left(\bar\psi^{(\mu,j)}_{s}\gamma^{\nu}\psi+\bar\psi\gamma^{\nu}\psi^{(\mu,j)}_{s}\right)\Biggr)
+\\ \nonumber
+\sum\limits_{j=1}^{2}\int\limits_{\frac{m}{\sqrt{2}}}^{\infty}d\mu'\biggl(-\frac{1}{4}f_{\rho\nu}^{(\mu',j)}
f^{\rho\nu}_{(\mu',j)}+\frac{{\mu'}^{2}}{2}a_{\rho}^{(\mu',j)}
a^{\rho}_{(\mu',j)}\biggr)+L_{int} \biggr]
\end{eqnarray}
where
$\psi^{(\mu,j)}_{s}=\cos(\theta_{(\mu)})\psi^{(\mu,j)}_{1}+\sin(\theta_{(\mu)})\gamma^{5}\psi^{(\mu,j)}_{2}$,
$e_{4}=\frac{\sqrt{\lambda}}{2^{\frac{3}{4}}\xi\sqrt{m}}\,e$,
$\tilde\mu=\sqrt{\mu^{2}-M^{2}}$,
$\tan(2\theta_{(\mu)})=\frac{\sqrt{\mu^{2}-M^{2}}}{M}$,
$\psi=\psi_{L}+\psi_{R}$ and $a_{\nu}$ are the lowest modes of the
fermion and vector boson fields (these fields can be identified
with the electron and the photon respectively), and $L_{int}$
stands for the term describing the interaction of the modes only
from the continuous spectra of the vector bosons and fermions. The
constant $C_{\psi,a}^{(\mu)}$ describes the coupling of the
massive fermions and vector bosons to the zero mode fermion, it
will not be used below, but one can find its definition in the
Appendix. Note, that there are no terms describing the interaction
of two zero-mode fields with one excited mode from the continuous
spectra, say, of the form $\sim
a_{\nu}^{(\mu,j)}\bar\psi\gamma^{\nu}\psi$. The corresponding
overlap integrals in such a case take the form
\begin{equation}
I_{(\mu,j)}\sim\int_{-\infty}^{\infty}f_{(\mu,j)}(z)\frac{1}{\cosh\left(\frac{m}{\sqrt{2}}\right)}dz
\end{equation}
or
\begin{equation}
I_{(\mu,j)}\sim\int_{-\infty}^{\infty}y_{(\mu,j)}(z)\frac{1}{\cosh\left(\frac{m}{\sqrt{2}}\right)}dz.
\end{equation}
Using (\ref{eqauxexc}) or (\ref{eqscal4}) it is not difficult to
show that $I_{(\mu,j)}\equiv 0$.

Now we are ready to consider the consequences of the model at
hand. Of course, one can quantize the resulting four-dimensional
effective theory (in this case the corresponding commutation and
anticommutation relations contain additional factor
$\delta(\mu-\mu')$) and then perform a calculation of particular
processes in such a theory. This is quite a bulky way, but we can
display the problems, arising in such a scenario, in a simpler
alternative way. To this end, let us pass from the continuous
spectra to discrete spectra replacing $\int
d\mu\to\sum\limits_{n}\Delta\mu$. We can perform the calculation
of particular processes in such a discretized theory and then take
the limit $\Delta\mu\to 0$ (of course, such a procedure does not
differ from putting the system into a box of a finite size). For
simplicity, we also introduce a cut-off scale $\tilde M$ such that
$$\int\limits_{\sqrt{\frac{m^2}{2}+M^{2}}}^{\infty}
d\mu\to\int\limits_{\sqrt{\frac{m^2}{2}+M^{2}}}^{\tilde M}d\mu.$$
In this case we get
\begin{eqnarray}
\int\limits_{\sqrt{\frac{m^2}{2}+M^{2}}}^{\tilde M}d\mu&\to
&\sum\limits_{n=1}^{N+1}\Delta\mu,\qquad \Delta \mu=\frac{\tilde
M-\sqrt{\frac{m^2}{2}+M^{2}}}{N},\\
\mu&\to &\mu_{n}=\sqrt{\frac{m^2}{2}+M^{2}}+(n-1)\Delta \mu+\epsilon\delta_{n1},\qquad\epsilon\ll\Delta\mu,\\
\sqrt{\Delta \mu}\,\psi^{(\mu_{n},j)}_{i}&\to & \psi^{n,j}_{i},\\
\sqrt{\frac{\mu_{n}\Delta
\mu}{\sqrt{\mu_{n}^{2}-M^{2}}}}\,a_{\nu}^{(\mu_{n},j)}&\to &
a_{\nu}^{n,j}.
\end{eqnarray}
The two latter transformations are necessary to acquire the
canonical normalization of four-dimensional fields. The extra
factor for the vector field arises because the redefinition
$\mu'\to\sqrt{\mu^2-M^2}$ in the last integral of (\ref{eqspact})
was made for simplicity.

The discretized action, following from (\ref{eqspact}), takes the
form
\begin{eqnarray}\label{eqspactdiscr}
S_{eff}=\int d^{4}x
\biggl[i\bar\psi\gamma^{\nu}\left(\partial_{\nu}-ie_{4}a_{\nu}\right)\psi-M\bar\psi\psi-\frac{1}{4}
f^{\rho\nu}f_{\rho\nu}+\\ \nonumber
+\sum\limits_{n=1}^{N+1}\sum\limits_{j=1}^{2}\Biggl(\sum\limits_{i=1}^{2}\left(i\bar\psi^{n,j}_{i}\gamma^{\nu}\left(\partial_{\nu}
-ie_{4}a_{\nu}\right)\psi^{n,j}_{i}
-\mu_{n}\bar\psi^{n,j}_{i}\psi^{n,j}_{i}\right)+
C_{\psi,a}^{n}a_{\nu}^{n,j}\left(\bar\psi^{n,j}_{s}\gamma^{\nu}\psi+\bar\psi\gamma^{\nu}\psi^{n,j}_{s}\right)-\\
\nonumber -\frac{1}{4}f_{\rho\nu}^{n,j}
f^{\rho\nu}_{n,j}+\frac{(\mu_{n}^{2}-M^{2})}{2}a_{\rho}^{n,j}
a^{\rho}_{n,j}\Biggr)+L_{int} \biggr].
\end{eqnarray}
Now we can proceed to particular processes in our model. We will
be interested in $\gamma\gamma\to\gamma\gamma$ scattering, where
$\gamma$ stands for the zero mode vector field $a_{\nu}$.
According to action (\ref{eqspactdiscr}), the corresponding
amplitude in the lowest order in the coupling constant is
schematically represented in Figure~1.
\begin{figure}[ht]
\begin{minipage}[h]{0.1\linewidth}
$A=$
\end{minipage}
\begin{minipage}[h]{0.25\linewidth}
\center{\includegraphics[width=0.99\linewidth]{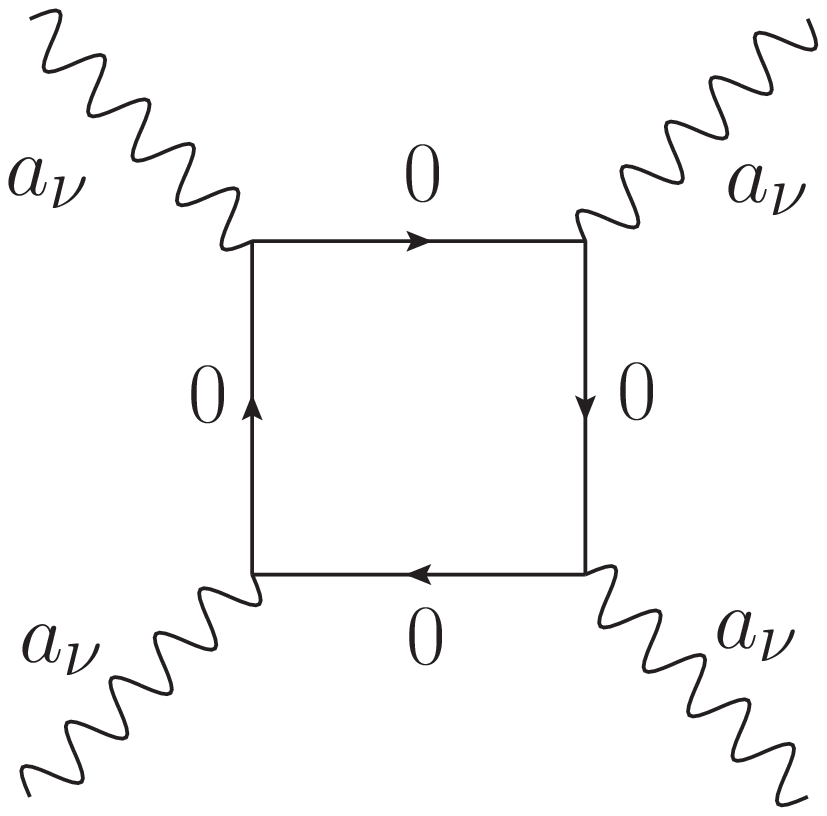}}
\end{minipage}
\begin{minipage}[h]{0.2\linewidth}
$$+{\sum\limits_{i=1,2}}\,{\sum\limits_{j=1,2}}\,{\sum\limits_{n\ne 0}}$$
\end{minipage}
\begin{minipage}[h]{0.25\linewidth}
\center{\includegraphics[width=0.99\linewidth]{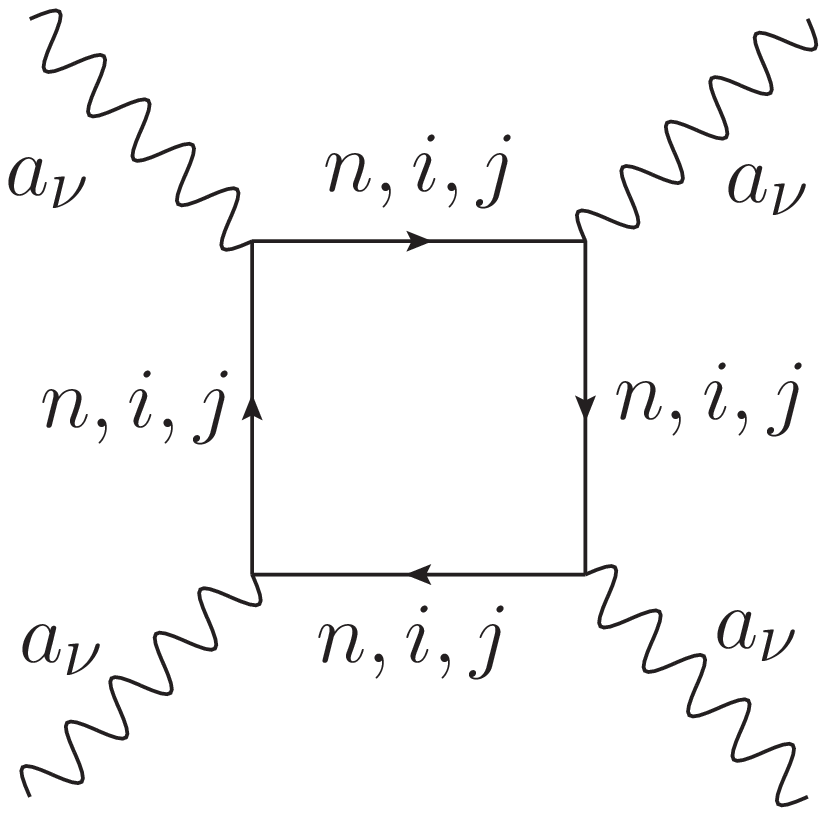}}
\end{minipage}
\caption{Schematic representation of the elastic $\gamma\gamma$
scattering amplitude. The number near the fermionic line
corresponds to the number of the massive fermionic mode which is
represented by this line.} \label{fig}
\end{figure}
In principle, in accordance with action (\ref{eqspactdiscr}), an
inelastic $\gamma\gamma$ scattering with massive vector bosons in
final states is also possible. However, below we will consider the
case when the energy of the initial photon in the c.m. frame is
much smaller than $\frac{m}{\sqrt{2}}$, in this case the creation
of massive vector bosons is kinematically forbidden. The amplitude
of the process presented in Figure~1 can be easily calculated
using the known results obtained in QED, see, for example,
\cite{AB}. For $\omega\ll\mu_{0}=M$, where $\omega$ is the energy
of the photon in the c.m. frame, the renormalized contribution of
each mode to the amplitude in the leading order in
$\frac{\omega}{\mu_{n}}$ is given by
\begin{equation}
A_{n}=\frac{\omega^4}{\mu_{n}^4}F(\theta),
\end{equation}
where the function $F(\theta)$ depends on the scattering angle
$\theta$ and the polarizations of the photons, $n=0$ corresponds
to the zero localized mode with $\mu_{0}=M$. The explicit form of
$F(\theta)$ will be irrelevant for our analysis, but it can be
found, for example, in \cite{AB}. Thus, we get
\begin{equation}\label{amplsum}
A=A_{0}+4\sum\limits_{n=1}^{N+1}A_{n}=\left(\frac{\omega^4}{\mu_{0}^4}+4\sum\limits_{n=1}^{N+1}\frac{\omega^4}{\mu_{n}^4}\right)
F(\theta).
\end{equation}
It is easy to show that this amplitude diverges in the limit
$\Delta\mu\to 0$. Indeed,
\begin{eqnarray}
A&=&\left(\frac{\omega^4}{\mu_{0}^4}+4\sum\limits_{n=1}^{N+1}\frac{\omega^4}{\mu_{n}^4}\right)F(\theta)
>4\sum\limits_{n=1}^{N+1}\frac{\omega^4}{\mu_{N+1}^4}F(\theta)=4(N+1)\frac{\omega^4}{\mu_{N+1}^4}F(\theta)=\\
\nonumber &=& 4\frac{\omega^4}{{\tilde M}^4}\left(1+\frac{\tilde
M-\sqrt{\frac{m^2}{2}+M^{2}}}{\Delta\mu}\right)F(\theta)\underset{\Delta\mu\to
0}{\longrightarrow}\infty,
\end{eqnarray}
leading to an infinite cross-section, even with the finite cut-off
scale $\tilde M$, which, of course, contradicts the experimental
data and predictions of the Standard Model. This infinity is not
the same as the infinities, which arise in QED and which can be
removed by applying the standard renormalization procedure, the
same growth of the amplitude would appear if we begun to add extra
"electrons" into the standard QED. Technically this happens
because the coupling constant of each fermion to the photon
remains finite in the limit $N\to\infty$. This differs from the
case of brane world models discussed in \cite{AK}, where there is
a single fermion in four-dimensional effective theory and a
continuous spectrum of massive vector modes, which couple to this
fermion (such a coupling is absent in the model discussed in this
paper because of our choice of the parameters). The interaction
term of the model discussed in \cite{AK} has the form
\begin{equation}
\int dm\Psi(m)\int d^{4}x B_{\mu}(x,m)\bar\psi\gamma^{\mu}\psi.
\end{equation}
After the discretization (with the redefinition of the field
$\sqrt{\Delta m}B_{\mu}(x,m_{n})\to B_{\mu}^n(x)$) we arrive at
\begin{equation}
\sum\limits_{n}\sqrt{\Delta m}\Psi_{n}\int d^{4}x
B_{\mu}^n(x)\bar\psi\gamma^{\mu}\psi.
\end{equation}
We see that in the model \cite{AK} the coupling constant tends to
zero in the limit $\Delta m\to 0$. Thus, the decreasing coupling
constants and increasing number of the modes compensate each other
and one gets a finite result in such a case, contrary to the case
of the model discussed in this paper.

Finally it should be noted that one can not obtain an analogous
result by taking, say, five-dimensional electrodynamics and
considering the scattering of five-dimensional photons with
$p_{4}=0$ holding for in- and out- photons, where $p_{4}$ is the
photon momentum in the extra coordinate in the c.m. frame. Let us
show it very schematically. The wave function of the photon in the
extra dimension in the case $p_{4}=0$ does not depend on $x^{4}$,
and integration over $x^{4}$ in the free photon action will lead
to an infinite value. Of course, this infinite value can be
incorporated into the photon field proper by making the
corresponding redefinition of this field, just by putting the
theory into the box of length $L$ in the extra dimension and then
redefining the field as $\sqrt{L}A_{\mu}\to A_{\mu}$. Formally the
effective action in this case appears to be analogous to the one
presented above (of course, without a mass gap etc.) The coupling
constant in this case appears to be redefined as
$e_{new}=e/\sqrt{L}$, which tends to zero at $L\to\infty$. In
order to have the coupling constant $e_{new}$ be finite in the
limit $L\to\infty$, which is exactly the case described by
(\ref{eqspact}), one should suppose that the initial coupling
constant in five-dimensional electrodynamics is infinite. Thus,
our effective theory described by action (\ref{eqspact}) is
similar in some sense to five-dimensional electrodynamics with an
infinite coupling constant.\footnote{The author is indebted to
M.V. Libanov for suggesting this analogy.}

\section{Passing from continuous to discrete spectrum -- does it help?}
In the previous section it was shown that the existence of a
continuous spectrum causes a problem in the low-energy effective
theory. Moreover, it looks as if such a problem is inherent to all
multidimensional models providing a continuous spectrum of excited
fermionic modes which have finite coupling constants to an
isolated vector zero mode. Of course, the simplest solution is
just to suppose that the cut-off scale $\tilde M$ is smaller than
$\sqrt{M^{2}+\frac{h^{2}}{\lambda}m^{2}}\approx\frac{hm}{\sqrt{\lambda}}$
for $M\ll \frac{hm}{\sqrt{\lambda}}$. But
$\frac{hm}{\sqrt{\lambda}}$ is exactly the factor which defines
the inverse width of the wave function of the zero fermionic mode,
see (\ref{wfrs}). Thus, the cut-off scale should be much larger
than $\frac{hm}{\sqrt{\lambda}}$, otherwise it appears to be very
unrealistic and rather strange to consider the cut-off scale to be
smaller than one of the typical scales of the effective theory we
are using to construct the model.

Another possible solution to the problem is to replace the
continuous spectrum by a completely discrete spectrum. This can be
done by changing the form of the interaction of five-dimensional
fermions with the background scalar field. Indeed, let us consider
the following form of the interaction term:
\begin{eqnarray}\label{interact2}
h\,\textrm{arctanh}\left({\frac{\sqrt{\lambda}}{m}\Phi}\right)\left(\bar\Psi_{2}\Psi_{2}-\bar\Psi_{1}\Psi_{1}\right)
\end{eqnarray}
instead of
\begin{eqnarray}\label{interact3}
h\Phi\left(\bar\Psi_{2}\Psi_{2}-\bar\Psi_{1}\Psi_{1}\right).
\end{eqnarray}
Of course, this term looks very artificial and unnatural, but note
that this is just a toy model. The equation of motion for the wave
function $f(z)$ (analogous to eq. (\ref{eqauxexc})) in this case
and with (\ref{kink}) takes the form
\begin{equation}\label{harmeq1}
\left(\mu^2-M^{2}+\frac{hm}{\sqrt{2}}\right)f(z)=-f''(z)+
\frac{h^{2}m^{2}}{2}z^{2}f(z).
\end{equation}
This is the standard quantum mechanical equation for the harmonic
oscillator. According to the fact that solutions to this equation
should be normalizable (see (\ref{normlocm})), we get the
well-known spectrum
\begin{equation}
\mu_{n}^2=M^{2}+hm\sqrt{2}n,\qquad n=0,1,...,\infty.
\end{equation}
It is evident that there is no continuous spectrum in this case.
The wave function of the zero mode takes the form
\begin{equation}
f(z)\sim e^{-\frac{hm}{2\sqrt{2}}z^{2}},\qquad {\tilde f}(z)\equiv
0.
\end{equation}
We see that this wave function decreases much faster with
$z\to\infty$ in comparison with (\ref{wfrs}). As for amplitude
(\ref{amplsum}), in this case it takes the form
\begin{equation}\label{amplsum1}
A=\frac{\omega^4}{\mu_{0}^4}
F(\theta)+2\sum\limits_{n=1}^{\infty}\frac{\omega^4}{\mu_{n}^4}
F(\theta)\approx
\omega^4\left(\frac{1}{M^4}+\frac{\pi^{2}}{6h^{2}m^{2}}\right)
F(\theta)
\end{equation}
for $hm\gg M^{2}$, it is finite and the contribution of the
excited modes appears to be highly suppressed for $hm\gg M^{2}$.

The existence of the continuous spectrum of the vector modes seems
to be not so dangerous, see small discussion on this topic in the
previous section. Nevertheless, one can also replace the
continuous spectrum by a discrete spectrum by considering a
Lagrangian of the (again very artificial) form
\begin{equation}\label{interact4}
S=-\int\xi^{2}\exp{\left(-\beta^{2}\textrm{arctanh}^{2}
\left({\frac{\sqrt{\lambda}}{m}\Phi}\right)\right)}\frac{1}{4}F^{MN}F_{MN}d^{4}xdz.
\end{equation}
It is very easy to show that in this case and with (\ref{kink}) we
again get the equation for the harmonic oscillator instead of
(\ref{eqscal4}) (the weight function $\sim e^{-c^{2}z^{2}}$ for a
vector field, leading to such an equation for the wave functions
of its modes, was discussed in \cite{ST}), moreover, by adjusting
$\beta$ we can get the equation with the same r.h.s. as that in
(\ref{harmeq1}). In such a fine-tuned case again there are no
terms describing the interaction of two zero-mode fields with one
excited mode in the effective four-dimensional Lagrangian. If the
masses of the excited states are large, they will not affect the
four-dimensional low-energy effective theory. It is interesting to
note that such a mechanisms of localization in principle can
simulate a compact extra dimension: from the four-dimensional
point of view such a theory can be treated as a five-dimensional
theory with a compact extra dimension, and the corresponding
localized modes -- as the Kaluza-Klein modes arising from a
compact extra dimension.

Nevertheless, this is not the whole story. There is another
continuous spectrum in the theory under consideration -- the
spectrum of the scalar fluctuations above the kink solution. Since
these modes couple both to the vector bosons and to the fermions,
one can expect divergencies, analogous to those discussed earlier,
resulting from the contribution of the scalar modes from the
continuous spectrum of the fluctuations above the background kink
solution. For example, in the case of interaction (\ref{sccoup0})
the scalar fluctuations couple quadratically to the vector field
(through the term in front of the kinetic term in
(\ref{sccoup0})), and there arises an interaction term of the form
\begin{equation}\label{scalcouplamunu}
\frac{\lambda}{2^{\frac{3}{2}}m}\int\limits_{\sqrt{2}m}^{\infty}
d\mu\int\varphi_{(\mu)}^{2}f^{\rho\nu}f_{\rho\nu}d^{4}x,
\end{equation}
where $\varphi_{(\mu)}$ are the fluctuations above the kink
solution. Such an interaction again can give an infinite
contribution to the elastic $\gamma\gamma$ scattering amplitude.
Note that one should not expect a compensation of divergencies
produced by the fermionic modes and the scalar modes as it happens
in SUSY theories -- the coupling constant of the scalar modes does
not include the electric charge $e$, see (\ref{scalcouplamunu}).
Indeed, there is no extra symmetry which could provide such a
compensation. It is very probable that an analogous problem arises
in the case of (\ref{interact2}) and (\ref{interact4}), because
there are quadratic in $\varphi_{(\mu)}$ and higher power terms in
the expansion of (\ref{interact4}) in terms of $\varphi_{(\mu)}$.

\section{Conclusion}
In this paper we discussed a model, describing a localization of
fermions and gauge bosons on a domain wall. It is shown that the
mechanism of localization of gauge fields, which is based on the
ideas proposed in \cite{KT,ST,Barnaveli}, provides a constant wave
function of the zero mode in the extra dimension and ensures the
universality of charge. The latter is crucial for localizing
non-abelian gauge fields and constructing realistic models of
fields localization. We also discussed the mechanism of fermion
localization, based on the well-known Rubakov-Shaposhnikov
mechanism \cite{RS}, which allows one to get four-dimensional
massive fermions localized on a domain wall. All the key
ingredients of this mechanism were presented in the literature
earlier, see \cite{DRT,LT,AAGS,AAGS2}, but we examined its main
properties in the case of localization on a domain wall and found
some very interesting consequences.

Both mechanisms of localization provide nonzero mass gaps between
the lowest zero and the subsequent modes. This allows us to
construct a simple five-dimensional model, resulting in an
effective four-dimensional theory, whose zero-mode sector
represents ordinary spinor electrodynamics localized on a domain
wall of the standard kink profile. But it appears that the model
possesses a very peculiar property -- it leads to an infinite
(renormalized) amplitude of $\gamma\gamma\to\gamma\gamma$
scattering process, which is known to be finite in ordinary
four-dimensional QED. A possible solution to this problem could be
in considering other forms of interaction of the background scalar
field with matter fields, which could lead to discrete spectra of
the excited modes of matter fields instead of the continuous
spectra. But most probably the existence of the continuous
spectrum of the fluctuations above the kink background solution
again would lead to unremovable divergencies in the amplitudes and
the cross-sections. Thus, there arises a question about more
realistic mechanisms of matter fields localization on a domain
wall.

\section*{Acknowledgements}

The author is grateful to M.Z.~Iofa, M.V.~Libanov and
I.P.~Volobuev for valuable discussions. The work was supported by
grant of Russian Ministry of Education and Science NS-4142.2010.2
and grant MK-3977.2011.2 of the President of Russian Federation.
\href{http://jaxodraw.sourceforge.net/}{JaxoDraw} program package
\cite{JD} was used to draw the Feynman diagrams presented in
Fig.~1.

\section*{Appendix}
The interaction term, describing the coupling of the massive
fermions and vector bosons to the zero mode fermion can be
obtained by substituting (\ref{sanz1})-(\ref{sanz3}) into
(\ref{sact}) and has the form
\begin{eqnarray}\nonumber
&&\frac{e\sqrt{\lambda}}{2^{\frac{3}{4}}\sqrt{m}}\sum\limits_{j,j'=1}^{2}\int
dz
\int\limits_{\frac{m}{\sqrt{2}}}^{\infty}d\mu_{1}\int\limits_{\sqrt{\frac{m^2}{2}+M^{2}}}^{\infty}d\mu\,
y_{(\mu_{1},j')}(z)f_{(\mu,j)}(z)J^{\nu}_{(\mu,j)}a_{\nu}^{(\mu_{1},j')}=\frac{e\sqrt{\lambda}}{2^{\frac{3}{4}}\sqrt{m}}\times
\\
\nonumber &&\times \sum\limits_{j,j'=1}^{2}\int dz
\int\limits_{\sqrt{\frac{m^2}{2}+M^{2}}}^{\infty}\frac{\mu'd\mu'}{\sqrt{{\mu'}^{2}-M^{2}}}\int\limits_{\sqrt{\frac{m^2}{2}+M^{2}}}^{\infty}d\mu\,
y_{(\sqrt{{\mu'}^{2}-M^{2}},j')}(z)f_{(\mu,j)}(z)J^{\nu}_{(\mu,j)}a_{\nu}^{(\sqrt{{\mu'}^{2}-M^{2}},j')}.
\end{eqnarray}
where
$J^{\nu}_{(\mu,j)}=\bar\psi^{(\mu,j)}_{s}\gamma^{\nu}\psi+\bar\psi\gamma^{\nu}\psi^{(\mu,j)}_{s}$.
It is evident from (\ref{eqauxexc}) (with
$\frac{h}{\sqrt{\lambda}}=\frac{1}{\sqrt{2}}$) and (\ref{eqscal4})
that
$f_{(\mu,j)}(z)=C_{f}^{(\mu)}y_{(\sqrt{{\mu}^{2}-M^{2}},j)}(z)$.
The constant $C_{f}^{(\mu)}$ simply reflects the fact that the
wave functions $y_{(\mu',j)}(z)$ and $f_{(\mu,j)}(z)$ possess
different normalization conditions. In general, one can calculate
$C_{f}^{(\mu)}$ using the exact forms of $y_{(\mu',j)}(z)$ and
$f_{(\mu,j)}(z)$ (for the case
$\frac{h}{\sqrt{\lambda}}=\frac{1}{\sqrt{2}}$ the wave functions
$\tilde
f_{(\mu,1)}(z)=c\sin\left(\sqrt{{\mu}^{2}-M^{2}-\frac{m^{2}}{2}}\,z\right)$,
$\tilde
f_{(\mu,2)}(z)=c\cos\left(\sqrt{{\mu}^{2}-M^{2}-\frac{m^{2}}{2}}\,z\right)$
and explicit form of the function $f_{(\mu,j)}(z)$ (and of the
function $y_{(\mu',j)}(z)$ also) can be obtained with the help of
(\ref{e5})).

Thus, we get
\begin{eqnarray}\nonumber
\frac{e\sqrt{\lambda}}{2^{\frac{3}{4}}\sqrt{m}}\sum\limits_{j,j'=1}^{2}\int
dz
\int\limits_{\sqrt{\frac{m^2}{2}+M^{2}}}^{\infty}\frac{\mu'd\mu'}{\sqrt{{\mu'}^{2}-M^{2}}}\int\limits_{\sqrt{\frac{m^2}{2}+M^{2}}}^{\infty}d\mu\,
y_{(\sqrt{{\mu'}^{2}-M^{2}},j')}(z)C_{f}^{(\mu)}y_{(\sqrt{{\mu}^{2}-M^{2}},j)}(z)\times\\
\nonumber\times
J^{\nu}_{(\mu,j)}a_{\nu}^{(\sqrt{{\mu'}^{2}-M^{2}},j')}=
\end{eqnarray}
\begin{eqnarray}\nonumber
=\frac{e\sqrt{\lambda}}{2^{\frac{3}{4}}\sqrt{m}\xi^{2}}\sum\limits_{j=1}^{2}
\int\limits_{\sqrt{\frac{m^2}{2}+M^{2}}}^{\infty}\frac{\mu'd\mu'}{\sqrt{{\mu'}^{2}-M^{2}}}\int\limits_{\sqrt{\frac{m^2}{2}+M^{2}}}^{\infty}d\mu\,
\delta\left(\sqrt{{\mu'}^{2}-M^{2}}-\sqrt{{\mu}^{2}-M^{2}}\right)\times\\
\nonumber\times
C_{f}^{(\mu)}J^{\nu}_{(\mu,j)}a_{\nu}^{(\sqrt{{\mu'}^{2}-M^{2}},j)}=
\frac{e\sqrt{\lambda}}{2^{\frac{3}{4}}\sqrt{m}\xi^{2}}
\sum\limits_{j=1}^{2}\int\limits_{\sqrt{\frac{m^2}{2}+M^{2}}}^{\infty}d\mu\,
C_{f}^{(\mu)}J^{\nu}_{(\mu,j)}a_{\nu}^{(\sqrt{{\mu}^{2}-M^{2}},j)}
\end{eqnarray}
and
$C_{\psi,a}^{(\mu)}=\frac{e\sqrt{\lambda}}{2^{\frac{3}{4}}\sqrt{m}\xi^{2}}C_{f}^{(\mu)}=\frac{e_{4}}{\xi}C_{f}^{(\mu)}$.


\begin{thebibliography}{20}
\bibitem{RS}
V.A.~Rubakov and M.E.~Shaposhnikov, Phys.\ Lett.\  B {\bf 125}
(1983) 136.

\bibitem{DS}
G.R.~Dvali and M.A.~Shifman, Phys. Lett. B {\bf 396} (1997) 64
[Erratum-ibid. B {\bf 407} (1997) 452].

\bibitem{DR}
S.L.~Dubovsky and V.A.~Rubakov,
  Int.\ J.\ Mod.\ Phys.\  A {\bf 16} (2001) 4331.

\bibitem{Rubakov}
V.A. Rubakov, Phys. Usp. {\bf 44} (2001) 871.

\bibitem{DRT0}
S.L.~Dubovsky, V.A.~Rubakov and P.G.~Tinyakov, JHEP {\bf 0008}
(2000) 041.

\bibitem{KT}
A.~Kehagias and K.~Tamvakis, Phys.\ Lett.\  B {\bf 504} (2001) 38.

\bibitem{ST}
M.E.~Shaposhnikov and P.~Tinyakov, Phys.\ Lett.\  B {\bf 515}
(2001) 442.

\bibitem{Barnaveli}
A.T.~Barnaveli and O.V.~Kancheli, Sov.\ J.\ Nucl.\ Phys.\  {\bf
52} (1990) 576. 

\bibitem{DRT}
S.L.~Dubovsky, V.A.~Rubakov and P.G.~Tinyakov, Phys.\ Rev.\  D
{\bf 62} (2000) 105011.

\bibitem{LT}
M.V.~Libanov and S.V.~Troitsky, Nucl.\ Phys.\  B {\bf 599} (2001)
319.

\bibitem{CG}
R.~Casadio and A.~Gruppuso, Phys.\ Rev.\  D {\bf 64} (2001)
025020.

\bibitem{AAGS}
A.A.~Andrianov, V.A.~Andrianov, P.~Giacconi and R.~Soldati, JHEP
{\bf 0307} (2003) 063.

\bibitem{AAGS2}
A.A.~Andrianov, V.A.~Andrianov, P.~Giacconi and R.~Soldati, JHEP
{\bf 0507} (2005) 003.

\bibitem{CHH}
A.E.R.~Chumbes, J.M. Hoff~da~Silva and M.B.~Hott, arXiv:1108.3821
[hep-th].

\bibitem{Flugge}
S. Fl\"ugge, {\em Practical Quantum Mechanics -- Volume 1 and
Volume 2}, 1971 (Springer).

\bibitem{UED}
C.~Macesanu, Int. J. Mod. Phys. A {\bf 21} (2006) 2259.

\bibitem{Barnaveli0}
A.T.~Barnaveli and O.V.~Kancheli, Sov.\ J.\ Nucl.\ Phys.\  {\bf
51} (1990) 573.

\bibitem{AB}
A.I. Ahiezer, V.B. Beresteckiy, {\em Quantum Electrodynamics},
1981 (Nauka, Moscow).

\bibitem{AK}
D.I.~Astakhov, D.V.~Kirpichnikov, Phys.\ Rev.\  {\bf D83 } (2011)
104031.

\bibitem{JD}
D.~Binosi, L.~Theussl, Comput.\ Phys.\ Commun.\  {\bf 161} (2004)
76.\\
D.~Binosi, J.~Collins, C.~Kaufhold, L.~Theussl, Comput.\ Phys.\
Commun.\  {\bf 180} (2009) 1709.

\end{thebibliography}
\end{document}